\documentclass[11pt,a4paper]{article}

\usepackage{jheppub}
\usepackage{epsfig}

\usepackage{amssymb,amsmath}
\usepackage{amsthm}
\usepackage{amsbsy}

\title{Higher-order renormalization of graphene many-body theory}

\author{J. Gonz\'alez}

\affiliation{Instituto de Estructura de la Materia,\\
        Consejo Superior de Investigaciones Cient\'{\i}ficas,\\ Serrano 123,
        28006 Madrid, Spain}

\emailAdd{gonzalez@iem.cfmac.csic.es}

\abstract{We study the many-body theory of graphene Dirac quasiparticles 
interacting via the long-range Coulomb potential, taking as a starting point 
the ladder approximation to different vertex functions. We test in this 
way the low-energy behavior of the electron system beyond the simple 
logarithmic dependence of electronic correlators on the high-energy cutoff, 
which is characteristic of the large-$N$ approximation. We show that 
the graphene many-body theory is perfectly renormalizable in the ladder 
approximation, as all higher powers in the cutoff dependence can be absorbed into 
the redefinition of a finite number of parameters (namely, the Fermi velocity 
and the weight of the fields) that remain free of infrared divergences even 
at the charge neutrality point. We illustrate this fact in the case of 
the vertex for the current density, where a complete cancellation 
between the cutoff dependences of vertex and electron self-energy corrections
becomes crucial for the preservation of the gauge invariance of the theory. 
The other potentially divergent vertex 
corresponds to the staggered (sublattice odd) charge density, which is made 
cutoff independent by a redefinition in the scale of the density operator. 
This allows to compute a well-defined, scale invariant anomalous dimension to
all orders in the ladder series, which becomes singular at a value of the 
interaction strength marking the onset of chiral symmetry breaking (and gap 
opening) in the Dirac field theory. The critical coupling we obtain in this
way matches with great accuracy the value found with a quite different 
method, based on the resolution of the gap equation, thus reassuring the 
predictability of our renormalization approach.}


\notoc
\begin{document}

\maketitle

\section{Introduction}

The discovery of graphene, the two-dimensional material made of a 
one-atom-thick carbon layer\cite{novo}, has opened new possibilities to 
investigate fundamental physics as well for devising technological 
applications. The electron system has relativistic-like invariance at 
low-energies, mimicking the behavior of Dirac fermions in two spatial 
dimensions\cite{geim,kim,rmp}. Moreover, the Coulomb repulsion between 
electrons constitutes the dominant interaction in the graphehe layer. This 
makes the low-energy theory to be a variant of Quantum Electrodynamics, but 
placed in the strong coupling regime as the ratio of $e^2$ to the Fermi 
velocity $v_F$ of the electrons is nominally larger than one.

There have been already several proposals to observe unconventional signatures 
of the interacting electrons in graphene. It has been for instance 
remarked that the interaction with impurities carrying a sufficiently large 
charge should result in anomalous screening properties of the graphene 
system\cite{nil,fog,shy,ter}. 
Furthermore, it was also found long ago that the own $e$-$e$ interaction in the 
layer should lead to a linear dependence on energy of the quasiparticle decay 
rate\cite{qlt}, as a consequence of the vanishing density of states at the 
charge neutrality point, and in agreement with measurements carried out in 
graphite\cite{exp}.

More precisely, it has been shown that the graphene electron system has the 
properties of a renormalizable quantum field theory, where the parameters flow 
with the energy scale\cite{np2}. In this framework, it was implied that the 
Fermi velocity should scale logarithmically towards larger values in the 
low-energy limit, what appears to be confirmed by recent experimental 
observations in graphene\cite{exp2}.

The graphene electron system is actually an example of electron liquid with 
strong many-body corrections, which depend significantly on the energy of the 
interaction processes. In practice, this is manifest in the logarithmic 
dependence on the high-energy cutoff needed to regularize the contributions to 
different quantities like the Fermi velocity or the weight of electron 
quasiparticles. In this type of electron liquid, one has to make sure that 
these divergences amount to the redefinition of a finite number of parameters 
in the system. In the context of quantum field theory, this property of 
renormalizability is crucial to guarantee the predictability of the theory
as quantum corrections are taken into account. Otherwise, there is the 
possibility that the singular dependences on the cutoff cannot be absorbed into 
the redefinition of a finite number of local operators of the bare theory. This 
may happen when they take for instance the form of momentum-dependent 
$\log (|{\bf p}|)$ corrections to local operators in the effective action, being 
then the reflection that the effective low-energy theory cannot be captured in 
terms of the local fields present in the original model. 

At this point, the best evidence of the renormalizability of the model of Dirac
fermions in graphene comes from the study of the theory in the limit of large 
number $N$ of fermion flavors\cite{prbr}, equivalent to the random-phase 
approximation (RPA). In this regime, it has been shown that 
all the cutoff dependences of the theory can be absorbed into redefinitions of 
the Fermi velocity and the weight of the electron quasiparticles, to all orders 
of the perturbative expansion in $e^2/v_F$ \cite{prbr} (for other studies of the 
$1/N$ expansion in graphene, see also Refs. \cite{ale} and \cite{son}). Anyhow, 
many-body corrections only exhibit at large $N$ a simple logarithmic dependence 
on the energy cutoff, which makes rather straightforward the renormalization of 
the theory at this stage.

In this paper we adopt an approach that is opposite in many aspects to that of 
the large-$N$ expansion, and that is able to probe the structure of the 
many-body corrections with arbitrary large powers of the cutoff dependence. That 
is based on the sum of the series of ladder diagrams, that we apply to different 
interaction vertices of the theory. Within this approach, we will be able to 
show that the divergent dependences on the cutoff can be reabsorbed in a finite 
number of parameters of the interacting theory, including the renormalization of 
the scale of different bilinears of the Dirac fermion fields. These renormalized 
quantities will prove to be independent of any infrared scale (Fermi energy, 
external momenta), making the low-energy limit of the many-body theory perfectly 
well-defined even at the charge neutrality point. 

From a practical point of view, the motivation for focusing on the sum 
of the ladder series lies in that it encodes the most divergent diagrams at each 
level of the perturbative expansion for the undoped electron system\cite{mis}. 
This makes highly nontrivial the process of renormalization, by which one has to 
remove in general divergent corrections that behave like the $n$-th power of the 
logarithm of the cutoff, when looking at the $n$-th perturbative level. 
In practice, we will illustrate the usefulness of the renormalization approach 
in the computation of observables like the anomalous dimensions of composite 
operators, which become determined just by the value of the renormalized 
coupling constant. This will allow us to address in particular the question of 
the dynamical breakdown of the chiral symmetry in the electron 
system\cite{khves,gus,vafek,khves2,her,jur,drut1,drut2,hands,hands2,gama,fer,ggg}, which can
be characterized in terms of the singular behavior of the corresponding 
anomalous dimension at a certain critical value of the coupling 
constant\cite{me,prb}.

\section{Dirac many-body theory}

Graphene is a 2D crystal of carbon atoms forming a honeycomb lattice, such that 
its low-energy electron quasiparticles are disposed into conical conduction and 
valence bands that touch at the six corners of the Brillouin zone\cite{rmp}. 
Of all six Fermi points, there are only two independent classes of electronic 
excitations. Thus, the low-energy electronic states can be encoded into a set of 
four-dimensional Dirac spinors $\{ \psi_i \}$,  which are characterized 
by having linear energy-momentum dispersion $\varepsilon ({\bf p}) = v_F |{\bf p}|$.
The index $i$ accounts for the two spin degrees of freedom, but may also 
allow to extend formally the analysis for a higher number $N$ of Dirac spinors.
The kinetic term of the hamiltonian in this low-energy theory is given by 
\begin{equation}
H_0 = - i v_F \int d^2 r \; \overline{\psi}_i({\bf r}) 
 \boldsymbol{\gamma}   \cdot \boldsymbol{\nabla}  \psi_i ({\bf r}) 
\label{h0} 
\end{equation}
where $\overline{\psi}_i = \psi_i^{\dagger} \gamma_0 $ and $\{ \gamma_{\sigma} \}$
is a collection of four-dimensional matrices such that 
$\{ \gamma_\mu, \gamma_\nu \} = 2 \: {\rm diag } (1,-1,-1)$. They can be 
conveniently represented in terms of Pauli matrices as
$\gamma_{0,1,2} = (\sigma_3, \sigma_3 \sigma_1, \sigma_3 \sigma_2) \otimes
 \sigma_3$, where the first factor acts on the two sublattice components of 
the honeycomb lattice and the second factor operates on the set of two independent
Fermi points.

In this paper we focus on the effects of the long-range Coulomb interaction in 
the graphene electron system. The density of states vanishes at the Fermi points 
connecting the conduction and valence bands, so that a sensible starting point for
the $e$-$e$ interaction is given by the unscreened potential 
$V({\bf r}) = e^2/4\pi |{\bf r}|$. The long-range Coulomb repulsion governs actually the 
properties of the electron system at low energies, since it is the
only interaction that is not suppressed, at the classical level, when scaling the many-body
theory in the limit of very large distances. If we add to (\ref{h0}) the contribution from the 
Coulomb interaction, we get the expression of the full hamiltonian
\begin{eqnarray}
H   =    - i v_F \int d^2 r \; \overline{\psi}_i({\bf r}) 
 \mbox{\boldmath $\gamma   \cdot \nabla $} \psi_i ({\bf r}) 
         + \frac{e^2}{8 \pi} \int d^2 r_1
\int d^2 r_2 \; \rho ({\bf r}_1) 
       \frac{1}{|{\bf r}_1 - {\bf r}_2|} \rho ({\bf r}_2)  \;\;\;\;\;
\label{ham}
\end{eqnarray}
with $\rho ({\bf r}) = \overline{\psi}_i ({\bf r}) \gamma_0 \psi_i ({\bf r})$. The total
action of the system is 
\begin{eqnarray}
S  & = &    \int dt  \int d^2 r \; \overline{\psi}_i({\bf r}) (i\gamma_0 \partial_t  
 + i v_F \mbox{\boldmath $\gamma   \cdot \nabla $} ) \psi_i ({\bf r})    \nonumber \\
  &   &     - \frac{e^2}{8 \pi} \int dt  \int d^2 r_1
\int d^2 r_2 \; \rho ({\bf r}_1) 
       \frac{1}{|{\bf r}_1 - {\bf r}_2|} \rho ({\bf r}_2)  \;\;\;\;\;
\label{s}
\end{eqnarray}
This action is invariant under the combined 
transformation of the space and time variables and the scale of the fields
\begin{equation}
t' = s t  \;\; , \;\;  {\bf r}' = s {\bf r}  \;\; , \;\; \psi_i' = s^{-1} \psi_i 
\label{scale}
\end{equation}
This means in particular that the strength of the 
interaction is not diminished (contrary to the case of a short-range interaction)
when zooming into the low-energy limit $s \rightarrow \infty $.

This analysis shows that the Coulomb repulsion mediated by the long-distance
$1/|{\bf r}|$ potential is the only interaction that may prevail 
in the low-energy regime of the electron system. It is clear that any other $e$-$e$ 
interaction without the $1/|{\bf r}|$ tail, as those that arise effectively
from phonon exchange, will be suppressed at least by a power of 
$1/s$ under the change of variables (\ref{scale}). The Dirac field theory 
with long-range Coulomb interaction has indeed the property of being 
scale invariant at this classical level, which provides a good starting point to
investigate the behavior of the many-body corrections upon scale 
transformations towards the long-wavelength limit  $s \rightarrow \infty $.

In fact, the many-body theory does not preserve in general the scale invariance 
of the classical action (\ref{s}), as a high-energy cutoff $\Lambda $ has to be 
introduced to obtain finite results in the computation of many-body 
corrections to different observables. The analysis of the cutoff dependence of the 
many-body theory provides deeper insight into the effective low-energy 
theory. If the theory is renormalizable, it must be possible to absorb all
powers of the cutoff dependence
into a redefinition of the parameters in the action (\ref{s}). This 
should be therefore modified to read
\begin{eqnarray}
\lefteqn{S =  Z_{\rm kin } \int dt  \int d^2 r \; \overline{\psi}_i({\bf r}) 
  (i\gamma_0 \partial_t  + i  Z_v \: v_F  \boldsymbol{\gamma}   
                        \cdot \boldsymbol{\nabla} ) \psi_i ({\bf r}) }   \nonumber \\
  &   &     - Z_{\rm int } \frac{e^2}{8 \pi} \int dt  \int d^2 r_1
\int d^2 r_2 \; \rho ({\bf r}_1) 
       \frac{1}{|{\bf r}_1 - {\bf r}_2|} \rho ({\bf r}_2)  \;\;\;\;\;
\label{sren}
\end{eqnarray}
The assumption is that $Z_{\rm kin }, Z_v$ and  $Z_{\rm int }$ (and other renormalization
factors for composite operators that do not appear in (\ref{s})) can only depend on the
cutoff, while they must be precisely chosen to render all electronic correlators
cutoff independent.  

The renormalizability of the graphene Dirac field theory is a nontrivial 
statement, since it amounts to the fact that that all the many-body corrections 
depending on the high-energy cutoff must reproduce the structure of the simple 
local operators that appear in (\ref{s}). It turns out for instance  
that many of the individual contributions to a given correlator have 
dependences in momentum space of the form $\log^n (|{\bf p}|) \;
\log^m (\Lambda)$. These are nonlocal corrections that cannot be reabsorbed
into the action (\ref{sren}), and the fact that all these nonlocal terms cancel
out in the final result for a correlator is a remarkable property of a 
renormalizable theory. Non-renormalizable theories have in this regard an
essential lack of predictability, as  $\log^n (|{\bf p}|)$ corrections do not 
make viable the characterization of the low-energy effective theory in terms of 
a few local operators, which may be in turn the reflection that it is not 
actually captured by the original fields formulated in the many-body theory.

\section{Electron self-energy and Fermi velocity renormalization}

\begin{figure}

\vspace{0.5cm}

\begin{center}
\mbox{\epsfxsize 12cm \epsfbox{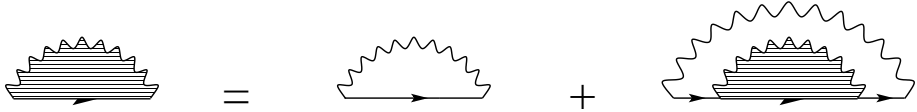}} 

\end{center}
\caption{Diagrammatic representation of the ladder approximation for the 
electron self-energy.}
\label{one}
\end{figure}

We first consider the cutoff dependence of the electron self-energy in the 
ladder approximation. We define this approach in terms of the self-consistent 
equation represented in Fig. \ref{one}. Diagrammatically, it corresponds to 
build the electron self-energy by iteration in the number of ^^ ^^ rainbow-like"
interactions between the Dirac fermion lines. A similar approach will be used 
afterwards to define the ladder approximation for the vertices of the charge 
and current density operators.

Before dealing with the actual ladder series, we establish our representation
of the free propagators by describing the computation of the lowest-order 
self-energy diagram. The free propagation of the Dirac fermions corresponds for 
instance to the expectation value
\begin{eqnarray}
\langle \psi_i ({\bf k}, \omega ) 
    \overline{\psi}_i ({\bf k}, \omega ) \rangle_{\rm free}
 &  =  &    i G_0 ({\bf k}, \omega )                          \nonumber       \\  
  &  =  &  i \frac{-\gamma_0 \omega  + v_F \boldsymbol{\gamma} \cdot {\bf k} }
                      {-\omega^2  + v_F^2 {\bf k}^2 - i\eta }     
\end{eqnarray}   
On the other hand, the interaction lines stand in momentum space for the 
product of $-i$ times the Fourier transform of the Coulomb potential, that 
turns out to be in two spatial dimensions
\begin{equation}
V({\bf q}) =  \frac{2\pi }{|{\bf q}|}
\end{equation}

The first-order electron self-energy diagram, that we will denote by 
$\Sigma_1 ({\bf k})$, needs to be regularized by introducing a high-energy
cutoff $\Lambda $ in the momentum integrals. We have actually
\begin{equation}
i \Sigma_1 ({\bf k}) = - \frac{e^2}{2}  
     \int \frac{d^2 p}{(2\pi)^2}  \int \frac{d\omega_p}{2\pi }  \gamma_0
   \frac{-\gamma_0 \omega_p  + v_F \boldsymbol{\gamma} \cdot {\bf p} }
                      {-\omega_p^ 2  + v_F^2 {\bf p}^2 - i\eta }  \gamma_0 
          \frac{1}{|{\bf k}-{\bf p}|}
\label{fo}
\end{equation}
which leads to a contribution proportional to $\boldsymbol{\gamma} \cdot {\bf k} $
that must be logarithmically divergent by simple dimensional counting. If we bound
the integration in momentum space such that $v_F |{\bf p}| < \Lambda $, we get
the result
\begin{equation}
 \Sigma_1 ({\bf k}) \approx  \frac{e^2}{16\pi } 
     \boldsymbol{\gamma} \cdot {\bf k}    \;  \log \Lambda  
\end{equation}
that corresponds to the well-known renormalization of the Fermi velocity by the 
Coulomb interaction in the Dirac many-body theory\cite{np2,prbr}.

From now on, we will choose a convenient regularization method to compute the 
divergent as well as the finite corrections to electronic correlators
at each perturbative level. That consists in 
the analytic continuation in the number of space dimensions\cite{ram}, by which 
the momentum integrals are computed at dimension $D = 2 - \epsilon $ \cite{np2}. With
this method, dependences on $\log \Lambda $ are traded in general by $1/\epsilon$
poles. In the above instance of the electron self-energy, we get after integration over $\omega_p$
\begin{equation}
i \Sigma_1 ({\bf k})     =     i \frac{e_0^2}{4}  \int \frac{d^D p}{(2\pi)^D}
  \boldsymbol{\gamma} \cdot {\bf p} \frac{1}{|{\bf p}|}  \frac{1}{|{\bf k}-{\bf p}|}
\end{equation}
where $e_0$ is a parameter whose dimensions are given by an auxiliary momentum 
scale $\mu $ through the relation
\begin{equation}
e_0 = \mu^{\epsilon/2} e
\end{equation} 
The calculation then proceeds as follows: 
\begin{eqnarray}
\Sigma_1 ({\bf k})   &  =  &  \frac{e_0^2}{4\pi}    
       \int \frac{d^D p}{(2\pi)^D}   \boldsymbol{\gamma} \cdot {\bf p} 
  \int_0^1 dx  \frac{x^{-1/2}  (1-x)^{-1/2}}{({\bf k}-{\bf p})^2 x + {\bf p}^2 (1-x)}
                                                               \nonumber                \\
   &  =  &  \frac{e_0^2}{4\pi}    
       \int \frac{d^D p}{(2\pi)^D}   \boldsymbol{\gamma} \cdot {\bf k} 
  \int_0^1 dx  \frac{x^{1/2}  (1-x)^{-1/2}}{{\bf p}^2  + {\bf k}^2 x(1-x)}
                                                                \nonumber                \\
    &  =  &  \frac{e_0^2}{4\pi}    \boldsymbol{\gamma} \cdot {\bf k} 
     \int_0^1 dx   \frac{\sqrt{x}}{\sqrt{1-x}}  \frac{\Gamma (1-D/2)}{(4\pi)^{D/2}}
                \frac{1}{ ({\bf k}^2 x(1-x))^{1-D/2} }               \nonumber          \\    
    &  =  &   \frac{e_0^2}{(4\pi)^2}    \boldsymbol{\gamma} \cdot {\bf k} 
         \frac{(4\pi)^{\epsilon/2}}{|{\bf k}|^{\epsilon}}
         \frac{\Gamma \left(\tfrac{1}{2}\epsilon\right) 
      \Gamma \left(\tfrac{3-\epsilon}{2}\right) \Gamma \left(\tfrac{1-\epsilon}{2}\right)}
                    {\Gamma (2-\epsilon )}
\label{loop}
\end{eqnarray}
From the latter expression we find the pole as $\epsilon \rightarrow 0$
\begin{equation}
 \Sigma_1 ({\bf k}) \approx  \frac{e^2}{16\pi } 
     \boldsymbol{\gamma} \cdot {\bf k}    \;  \frac{1}{\epsilon }  
\label{eps}
\end{equation}

We are anyhow interested in the result of computing the electron self-energy in the
ladder approximation defined in Fig. \ref{one}. It is
easily realized that the solution  $\Sigma_{\rm ladder} ({\bf k})$ of the 
self-consistent equation must have the structure
\begin{equation}
\Sigma_{\rm ladder} ({\bf k}) =  f ({\bf k}) \: \boldsymbol{\gamma} \cdot {\bf k}
\end{equation}
with a scalar function $f ({\bf k})$. The self-energy then satisfies
\begin{equation}
i\Sigma_{\rm ladder} ({\bf k}) =  i\Sigma_1 ({\bf k})    
        +   \frac{e_0^2}{2}    \int \frac{d^D p}{(2\pi)^D}   
          \int \frac{d\omega_p}{2\pi }     
\Sigma_{\rm ladder} ({\bf p}) 
    \frac{ \omega_p^2  + v_F^2 {\bf p}^2 }
                      {(-\omega_p^ 2  + v_F^2 {\bf p}^2 - i\eta)^2} 
  \frac{1}{|{\bf k}-{\bf p}|}              
\label{selfl}
\end{equation}

The solution of Eq. (\ref{selfl}) reflects a particular feature of 
the graphene many-body theory in the static limit (i.e. when the effective 
interaction is supposed to be frequency independent). It can be checked that the 
second term in the self-consistent equation identically vanishes, as a result of 
performing the integration over the frequency variable, and irrespective of the 
actual momentum dependence of $\Sigma_{\rm ladder} ({\bf k})$. We have
indeed, by performing a Wick rotation to imaginary frequency 
$\overline{\omega}_k = -i  \omega_k$,
\begin{equation}
     \int \frac{d\omega_p}{2\pi }  
        \frac{ \omega_p^2  + v_F^2 {\bf p}^2 }
                      {(-\omega_p^ 2  + v_F^2 {\bf p}^2 - i\eta)^2} 
  =  i \int \frac{d \overline{\omega}_p}{2\pi }  
    \frac{-\overline{\omega}_p^2  + v_F^2  {\bf p}^2 }
                      {(\overline{\omega }_p^ 2  + v_F^2 {\bf p}^2  )^2}
  =  0
\label{res}
\end{equation}
This result implies that the solution of Eq. (\ref{selfl}) must coincide with the 
first-order contribution
\begin{equation}
\Sigma_{\rm ladder} ({\bf k})     =   \Sigma_1 ({\bf k})
\end{equation}

\begin{figure}

\vspace{0.5cm}

\begin{center}
\mbox{\epsfxsize 12cm \epsfbox{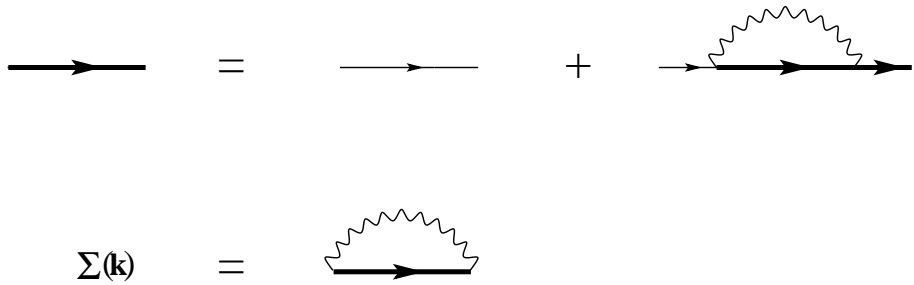}} 

\end{center}
\caption{Diagrammatic representation of the self-consistent exchange 
approximation for the electron self-energy.}
\label{two}
\end{figure}

This vanishing of higher-order corrections to the electron self-energy in the 
ladder approximation can be actually seen as the consequence of a wider
symmetry operating in the graphene many-body theory. We can extend the 
sum of self-energy diagrams in the ladder series to include contributions
where the electron lines in the ladder diagrams are corrected by the 
own electron self-energy. This leads to the sum of a much broader class of
diagrams, that are encoded in the self-consistent exchange 
approximation depicted in Fig. \ref{two}. If we represent the electron 
self-energy in this approach by  
\begin{equation}
\Sigma_{\rm SCEX} ({\bf k}) =  
     \tilde{f} ({\bf k}) \:  \boldsymbol{\gamma} \cdot {\bf k}  
\end{equation}
the self-consistent equation can be written as
\begin{equation}
 i \Sigma_{\rm SCEX} ({\bf k}) =   
          - \frac{e_0^2}{2}      
       \int \frac{d^D p}{(2\pi)^D}  \int \frac{d\omega_p}{2\pi }  
  \gamma_0  \frac{-\gamma_0 \omega_p  + (v_F + \tilde{f} ({\bf p})) \boldsymbol{\gamma} \cdot {\bf p} }
                      {-\omega_p^2  + (v_F + \tilde{f} ({\bf p}))^2 {\bf p}^2 - i\eta } \gamma_0
            \frac{1}{|{\bf k}-{\bf p}|}
\label{HF}
\end{equation}

The key observation is that, for the same reason that the final expression
for the first-order self-energy (\ref{fo}) does not depend on the Fermi velocity $v_F$,
the integral in Eq. (\ref{HF}) turns out to be independent of the function 
$\tilde{f} ({\bf k})$. One can for instance redefine the scale of the frequency 
from $\omega_p$ to $(1 +  \tilde{f} ({\bf p})/v_F )\omega_p$, in such a way 
that the integrand in Eq. (\ref{HF}) falls into the corresponding first-order expression
in (\ref{fo}). This proves that, also in the more comprehensive self-consistent 
exchange approximation, the electron self-energy coincides with the first-order 
result
\begin{equation}
\Sigma_{\rm SCEX} ({\bf k}) = \Sigma_1 ({\bf k}) 
\label{nr}
\end{equation}

In its simplicity, the result expressed in Eq. (\ref{nr}) accounts for the 
vanishing of a vast class of corrections to the electron self-energy in 
graphene. It can be interpreted as a kind of no-renormalization theorem that 
protects the Fermi velocity from being modified by higher-order effects, which 
remains valid under the assumption of static (frequency-independent) screening 
of the Coulomb interaction. As we will see, this translates into a remarkable 
cancellation of corrections to the vertex of the current density operator, in 
a nontrivial check of a gauge invariance that is hidden in the original 
formulation of the theory.

\section{Charge and current density correlations}

\begin{figure}

\vspace{0.5cm}

\begin{center}
\mbox{\epsfxsize 10cm \epsfbox{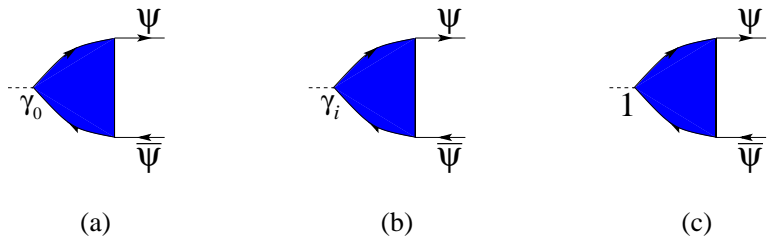}} 

\end{center}
\caption{Diagrammatic representation of the vertices for (a) total charge density,
(b) current density, and (c) staggered charge density.}
\label{three}
\end{figure}

We study next the way in which many-body corrections dress the different
interaction vertices in graphene. This includes the inspection of the own 
Coulomb interaction, that we analyze by looking for corrections of the coupling to 
the total charge density. The Dirac field
theory allows anyhow for the consideration of more general vertices that take 
into account the pseudospin current $\boldsymbol{\gamma}$ and the spinor structure
of the fermion fields. We will pay attention in what follows to the vertices
for the total charge, the pseudospin current, and the staggered (sublattice odd)
charge density, which are represented in Fig. \ref{three}.

The analysis of the vertices for the total charge and the pseudospin current is 
particularly relevant since, if we think of them as operators that may be 
switched on in the action of the electron system, it becomes clear that they 
should be related by gauge invariance to the terms in the kinetic action. That 
is, we can start with an extended action given by 
\begin{eqnarray}
\lefteqn{S =  Z_{\rm kin } \int dt  \int d^2 r \; \overline{\psi}_i({\bf r}) 
  (i\gamma_0 \partial_t  + i  Z_v \: v_F  \boldsymbol{\gamma}   
                        \cdot \boldsymbol{\nabla} ) \psi_i ({\bf r}) }   \nonumber \\
  &   &     +  e \int dt  \int d^2 r \; \overline{\psi}_i({\bf r}) 
  (Z_{\rm int }' \gamma_0  A_0  +  Z_{\rm int }''  \boldsymbol{\gamma}   
                        \cdot {\bf A} ) \psi_i ({\bf r}) 
\label{szz}
\end{eqnarray}
where $A_0$ and ${\bf A}$ play the role of auxiliary fields mediating the
interactions of the total charge and the pseudospin current. A gauge
transformation of the Dirac fields
\begin{equation}
\widetilde{\psi}_i ({\bf r}) = e^{ie \theta ({\bf r},t)} \psi_i ({\bf r})
\label{gauge}
\end{equation}
amounts to a shift of the auxiliary fields $A_0$ and ${\bf A}$. Thus,
the invariance of the many-body theory under (\ref{gauge}) can be tested
by checking that the renormalization factors $Z_{\rm int }'$ and
$Z_{\rm int }''$ match with the respective factors from the renormalization 
of the electron self-energy.

This question of the gauge invariance is an interesting point regarding the
graphene many-body theory, as it was shown long ago that the four-fermion
interactions in the graphene electron system can be obtained from a suitable
projection of the full relativistic interaction mediated by photons in three
spatial dimensions\cite{np2}. It has been actually proven that the 
renormalization of the theory, when carried out to first order in perturbation 
theory, is consistent with the above mentioned gauge invariance. In the 
present instance, we will also use the renormalization properties of the 
vertices to check the underlying gauge invariance to higher orders 
in the ladder approximation supplemented by electron self-energy corrections.

\begin{figure}

\vspace{0.5cm}

\begin{center}
\mbox{\epsfxsize 10cm \epsfbox{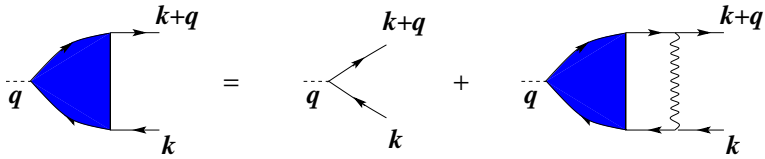}} 

\end{center}
\caption{Self-consistent diagrammatic equation for a generic vertex $\Gamma_i$
in the ladder approximation.}
\label{four}
\end{figure}

As in the case of the electron self-energy, we define the 
ladder approach for the vertices by means of a 
self-consistent equation, represented now in Fig. \ref{four}.
In principle, one can solve the equation by means of an iterative 
procedure, ending up with the equivalent of a ladder series for the different
vertices. We will then improve this diagrammatic approach in a second 
stage, by assuming that the internal fermion lines in the self-consistent 
equation are themselves corrected by the electron 
self-energy, which will prove to be crucial to preserve the gauge invariance 
of the theory.

\subsection{Charge density vertex}

We define the vertex for the total charge density in frequency and momentum 
space as 
\begin{equation}
 \Gamma_0 ({\bf q},\omega_q;{\bf k},\omega_k)  =
   \langle  \rho ({\bf q},\omega_q) 
        \psi_i ({\bf k}+{\bf q},\omega_k + \omega_q) 
           \overline{\psi}_i ({\bf k},\omega_k) \rangle_{\rm 1PI}
\label{gamma0}
\end{equation}
where $\rho $ is given in real space by
\begin{equation}
\rho ({\bf r}) = \overline{\psi}_i ({\bf r}) \gamma_0 \psi_i ({\bf r})
\end{equation}
and the right-hand-side of (\ref{gamma0})
is computed by considering only the one-particle-irreducible vertex 
diagrams. In this way, the possible renormalization required to render 
$\Gamma_0$ cutoff independent should amount to a simple multiplication of the
vertex by a factor, that is the same $Z_{\rm int}'$ appearing in (\ref{szz}).

The vertex $\Gamma_0$ is a dimensionless quantity, which means that,
in order to isolate the singular dependence on the cutoff, it is enough to study 
the limit ${\bf q} \rightarrow 0$ and $\omega_q \rightarrow 0$. 
Then, the self-consistent equation represented in Fig. \ref{four} becomes
\begin{equation}
\Gamma_0 ({\bf 0},0;{\bf k},\omega_k) = \gamma_0 +  
    i \frac{e^2_0}{2}  \int \frac{d^D p}{(2\pi )^D} \frac{d\omega_p}{2\pi } 
      \Gamma_0 ({\bf 0},0;{\bf p},\omega_p) 
      \frac{\omega_p^2  + v_F^2 {\bf p}^2 }
                 {(-\omega_p^2  + v_F^2 {\bf p}^2 - i\eta )^2} 
               \frac{1}{|{\bf k}-{\bf p}|}
\label{selfg0}
\end{equation}

It is clear that the solution of (\ref{selfg0}) cannot depend on the 
frequency $\omega_k$ of the external fermion lines. Therefore, 
the integral at the right-hand-side of the equation must be identically
zero, for the same reason that the integral in Eq. (\ref{selfl}) was also
vanishing. This means that the vertex $\Gamma_0$ is
independent of the cutoff in the ladder approximation. Repeating here the 
argument at the end of the last section, it turns out that the same 
statement holds true even when the fermion propagators in (\ref{selfg0})
are corrected with the electron self-energy (\ref{loop}). Again, the integral
in (\ref{selfg0}) vanishes irrespective of the momentum-dependent 
corrections to $v_F$, leaving $\Gamma_0$ cutoff independent in this 
approach.

The cutoff independence of $\Gamma_0$ agrees with the 
absence of wavefunction renormalization ($Z_{\rm kin} = 1$) in the self-consistent
exchange approximation applied to the electron self-energy. 
The trivial result 
\begin{equation}
Z_{\rm kin} = Z_{\rm int}' = 1
\label{zkin}
\end{equation}
is the first check of the gauge invariance of the theory. The vanishing of the 
many-body corrections lies in this instance in the particular structure of the ladder
approximation and, in this regard, it is a result that holds even after 
dressing the interaction with the static (frequency-independent) RPA  
screening of the Coulomb potential.

\subsection{Current density vertex}

The irreducible vertex for the current density is defined in this case by
\begin{equation}
 \boldsymbol{\Gamma}_{c} ({\bf q},\omega_q;{\bf k},\omega_k)  =
   \langle  \boldsymbol{\rho}_{c} ({\bf q},\omega_q) 
        \psi_i ({\bf k}+{\bf q},\omega_k + \omega_q) 
           \overline{\psi}_i ({\bf k},\omega_k) \rangle_{\rm 1PI}
\label{gammac}
\end{equation}
where the current density operator is given in real space by
\begin{equation}
\boldsymbol{\rho}_{c} ({\bf r}) = \overline{\psi}_i ({\bf r}) \: \boldsymbol{\gamma} \: \psi_i ({\bf r})
\end{equation}
We anticipate the fact that the computation of the vertex may give rise to 
dependences on the high-energy cutoff, that are supposed to be absorbed in the 
renormalization factor $Z_{\rm int}''$.

The vertex $\boldsymbol{\Gamma}_{c}$ is a two-dimensional vector,
but its analysis can be greatly simplified by considering as before the limit
${\bf q} \rightarrow 0$ and $\omega_q \rightarrow 0$. The self-consistent
equation depicted in Fig. \ref{four} takes then the form
\begin{eqnarray}
\lefteqn{\boldsymbol{\Gamma}_{c} ({\bf 0},0;{\bf k},\omega_k) = }
                                                    \nonumber       \\
  &  &     {\boldsymbol{\gamma}}     +  
    i \frac{e^2_0}{2} \int \frac{d^D p}{(2\pi )^D} \frac{d\omega_p}{2\pi } 
     \gamma_0 \frac{-\gamma_0 \omega_p  + v_F \boldsymbol{\gamma} \cdot {\bf p} }
                 {-\omega_p^2  + v_F^2 {\bf p}^2 - i\eta } 
     \boldsymbol{\Gamma}_{c} ({\bf 0},0;{\bf p},\omega_p) 
      \frac{-\gamma_0 \omega_p  + v_F \boldsymbol{\gamma} \cdot {\bf p} }
                 {-\omega_p^2  + v_F^2 {\bf p}^2 - i\eta } \gamma_0
               \frac{1}{|{\bf k}-{\bf p}|}   \;\;\;\;\;\;\;
\label{selfgc}
\end{eqnarray}

We resort at this point to an iterative resolution of (\ref{selfgc}), 
by which we can obtain a recursion between consecutive orders in 
the expansion of the vertex in powers of the interaction strength.
This procedure shows that 
$\boldsymbol{\Gamma}_{c}({\bf 0},0;{\bf k},\omega_k)$ has 
a part proportional to $\boldsymbol{\gamma}$ and another contribution 
proportional to ${\bf k} (\boldsymbol{\gamma} \cdot {\bf k})$. From
dimensional arguments, one can see that the solution
of (\ref{selfgc}) must take the form 
\begin{equation}
\boldsymbol{\Gamma}_{c}({\bf 0},0;{\bf k},\omega_k)  = 
    \boldsymbol{\gamma}
 \left(1 + \sum_{n=1}^{\infty} \lambda_0^n 
                           \frac{r_n }{|{\bf k}|^{n\epsilon}} \right)
   +   {\bf n}_{\bf k} (\boldsymbol{\gamma} \cdot {\bf n}_{\bf k})
            \sum_{n=1}^{\infty} \lambda_0^n 
                           \frac{r_n' }{|{\bf k}|^{n\epsilon}}
\label{exp}
\end{equation}
where we have called ${\bf n}_{\bf k} =  {\bf k}/|{\bf k}|$ and
\begin{equation}
\lambda_0 = \frac{e_0^2}{4\pi  v_F} 
\end{equation}
   
If we insert for instance a given order of the expansion with 
coefficient $r_n$ inside the integral in Eq. (\ref{selfgc}), we get
\begin{eqnarray}
\lefteqn{  - \frac{e^2_0}{2}  \int \frac{d^D p}{(2\pi )^D} \frac{d\overline{\omega}_p}{2\pi } 
     \gamma_0 \frac{-i \gamma_0 \overline{\omega}_p  + v_F \boldsymbol{\gamma} \cdot {\bf p} }
                 {\overline{\omega}_p^2  + v_F^2 {\bf p}^2 } 
       \boldsymbol{\gamma} \frac{r_n }{|{\bf p}|^{n\epsilon}}
      \frac{-i \gamma_0 \overline{\omega}_p  + v_F \boldsymbol{\gamma} \cdot {\bf p} }
                 {\overline{\omega}_p^2  + v_F^2 {\bf p}^2 } \gamma_0
               \frac{1}{|{\bf k}-{\bf p}|}  }     \nonumber    \\
  &  =  &  r_n \frac{e_0^2}{4v_F} \int \frac{d^D p}{(2\pi )^D}
        \left( \boldsymbol{\gamma}  \frac{1}{|{\bf p}|^{1+n\epsilon}} 
         -   {\bf p} (\boldsymbol{\gamma} \cdot {\bf p})  
              \frac{1}{|{\bf p}|^{3+n\epsilon}}  \right)
                  \frac{1}{|{\bf k}-{\bf p}|}         \nonumber    \\
  &  =  &  r_n \frac{e_0^2}{4v_F} \int \frac{d^D p}{(2\pi )^D}
 \left( \boldsymbol{\gamma} - {\bf n}_{\bf p} (\boldsymbol{\gamma} \cdot {\bf n}_{\bf p}) \right)
       \frac{\Gamma \left( 1 + \tfrac{n\epsilon}{2} \right)}
                  {\sqrt{\pi} \Gamma \left( \tfrac{1+n\epsilon}{2} \right)}
     \int_0^1  \frac{x^{-1/2} (1-x)^{-(1-n\epsilon)/2 }}
               {(({\bf k}-{\bf p})^2 x + {\bf p}^2 (1-x))^{1+n\epsilon/2}}              
                                                       \nonumber    \\
  &  =  &  \boldsymbol{\gamma} \: \lambda_0 \frac{r_n}{|{\bf k}|^{(n+1)\epsilon}} 
                  \frac{(4\pi)^{\epsilon/2}}{4} 
    \left( \frac{\Gamma \left(\tfrac{n+1}{2}\epsilon \right) 
           \Gamma \left(\tfrac{1-(n+1)\epsilon}{2} \right) 
           \Gamma \left(\tfrac{1-\epsilon}{2} \right)}
               {\sqrt{\pi} \Gamma \left(\tfrac{1+n\epsilon}{2} \right) 
                 \Gamma \left(1-\tfrac{(n+2)\epsilon}{2} \right) }        
    -  \frac{\Gamma \left(\tfrac{n+1}{2}\epsilon \right) 
           \Gamma \left(\tfrac{1-(n+1)\epsilon}{2} \right) 
           \Gamma \left(\tfrac{3-\epsilon}{2} \right)}
               {2\sqrt{\pi} \Gamma \left(\tfrac{3+n\epsilon}{2} \right) 
                 \Gamma \left(2-\tfrac{(n+2)\epsilon}{2} \right) }   \right)
                                                      \nonumber    \\
  &  &  -  {\bf n}_{\bf k} (\boldsymbol{\gamma} \cdot {\bf n}_{\bf k}) 
           \: \lambda_0 \frac{r_n}{|{\bf k}|^{(n+1)\epsilon}}      
              \frac{(4\pi)^{\epsilon/2}}{4} 
     \frac{\Gamma \left(1 + \tfrac{n+1}{2}\epsilon \right) 
           \Gamma \left(\tfrac{3-(n+1)\epsilon}{2} \right) 
           \Gamma \left(\tfrac{1-\epsilon}{2} \right)}
               {\sqrt{\pi} \Gamma \left(\tfrac{3+n\epsilon}{2} \right) 
                 \Gamma \left(2-\tfrac{(n+2)\epsilon}{2} \right) } 
\label{integ}
\end{eqnarray}
On the other hand, by inserting any term of the expansion (\ref{exp}) with 
$r_n'$ coefficient inside the integral of the self-consistent equation, we get
always a vanishing result due to Eq. (\ref{res}).
We obtain therefore the recurrence relations
\begin{equation}
r_{n+1}   =    \frac{(4\pi)^{\epsilon/2}}{4} 
    \left( \frac{\Gamma \left(\tfrac{n+1}{2}\epsilon \right) 
           \Gamma \left(\tfrac{1-(n+1)\epsilon}{2} \right) 
           \Gamma \left(\tfrac{1-\epsilon}{2} \right)}
               {\sqrt{\pi} \Gamma \left(\tfrac{1+n\epsilon}{2} \right) 
                 \Gamma \left(1-\tfrac{(n+2)\epsilon}{2} \right) }        
    -  \frac{\Gamma \left(\tfrac{n+1}{2}\epsilon \right) 
           \Gamma \left(\tfrac{1-(n+1)\epsilon}{2} \right) 
           \Gamma \left(\tfrac{3-\epsilon}{2} \right)}
               {2\sqrt{\pi} \Gamma \left(\tfrac{3+n\epsilon}{2} \right) 
            \Gamma \left(2-\tfrac{(n+2)\epsilon}{2} \right) }   \right)  r_n
                                                          \label{rec}       
\end{equation}
\begin{equation}
r_{n+1}'   =    -  \frac{(4\pi)^{\epsilon/2}}{4} 
     \frac{\Gamma \left(1 + \tfrac{n+1}{2}\epsilon \right) 
           \Gamma \left(\tfrac{3-(n+1)\epsilon}{2} \right) 
           \Gamma \left(\tfrac{1-\epsilon}{2} \right)}
               {\sqrt{\pi} \Gamma \left(\tfrac{3+n\epsilon}{2} \right) 
                 \Gamma \left(2-\tfrac{(n+2)\epsilon}{2} \right) }  r_n
\end{equation}

We observe from (\ref{rec}) that the expansion of the vertex develops 
increasing divergences in the cutoff, that manifest as poles in the limit
$\epsilon \rightarrow 0$. The key point is whether these divergences can 
be absorbed by a suitable renormalization factor. The current density 
$\boldsymbol{\rho}_{c} ({\bf r})$ is not an elementary field of the many-body
theory, which means that its correlators need to be renormalized by appropriate
rescaling of the own current density. Alternatively, if we include the 
composite field with its own coupling in the action (\ref{szz}), it 
is the renormalization factor $Z_{\rm int }''$ which needs to be adjusted 
to render the correlators cutoff independent. Then, 
a renormalized vertex $\boldsymbol{\Gamma}_{c, {\rm ren}}$, finite in the 
limit $\epsilon \rightarrow 0$, has to be obtained by the multiplicative 
renormalization
\begin{equation}
\boldsymbol{\Gamma}_{c, {\rm ren}} = Z_{\rm int }'' 
   \boldsymbol{\Gamma}_{c}  
\end{equation}

The renormalization factor may have in general the structure
\begin{equation}
Z_{\rm int }'' = 1 + \sum_{i=1}^{\infty} \frac{c_i (\lambda )}{\epsilon^i}
\label{poles}
\end{equation}
in terms of the dimensionless physical coupling
\begin{equation}
\lambda = \frac{e^2}{4\pi  v_F} 
\end{equation}
It is a nontrivial fact that all the poles in $\boldsymbol{\Gamma}_{c}$ 
may be canceled against multiplication by $Z_{\rm int }''$, allowing only for
the dependence of the coefficients $c_i$ on the coupling constant. We have
checked that this is indeed the case, up to the order $\lambda^{18}$ we have 
been able to carry out the numerical computation of the vertex. We have found 
for instance for the first terms in the expansion (\ref{poles})
\begin{eqnarray}
c_1 (\lambda ) & = &  -\frac{1}{4} \lambda -\frac{1}{64} (1+\log (16)) \lambda^2
    -\frac{1}{384} \left(-1+3 \log ^2(4)+\log (64)\right) \lambda^3   \nonumber   \\
    &  &   -\frac{-9-72 \log (2)+384 \log ^2(2)+128 \log ^2(2) \log (16)+12 \zeta (3)}{24576} 
                      \lambda^4                                      \nonumber   \\
   &  & -\frac{3-30 \log (2)-60 \log ^2(2)+400 \log ^3(2)+400 \log ^4(2)+6 \zeta (3) 
   + 6 \log (16) \zeta (3)}{24576}  
                 \lambda^5                                           \nonumber     \\      
    &  &                                +  \ldots          \label{cs0}             \\
c_2 (\lambda ) & = &  \frac{1}{32} \lambda^2  +\frac{1}{256} (1+4 \log (2)) \lambda^3
          -\frac{13-120 \log (2)-120 \log (2) \log (4)}{24576}    \lambda^4     \nonumber  \\
   &  & -\frac{13+64 \log (2)-528 \log ^2(2)-176 \log ^2(2) \log (16)-12 \zeta (3)}{98304}  
                 \lambda^5     +  \ldots                                        \\
c_3 (\lambda ) & = &  -\frac{1}{384} \lambda^3  -\frac{1+4\log (2)}{2048}  \lambda^4
         -\frac{-5+72 \log (2)+144 \log ^2(2)}{98304}    \lambda^5   +  \ldots      \\
c_4 (\lambda ) & = &   \frac{1}{6144} \lambda^4 
      +\frac{1+4 \log (2)}{24576}   \lambda^5   +  \ldots                      \\
c_5 (\lambda ) & = &    -\frac{1}{122880}   \lambda^5   +   \ldots 
\label{cs}
\end{eqnarray}
The important point about this result for $Z_{\rm int }''$ is that it 
does not depend on the momenta of the vertex $\boldsymbol{\Gamma}_{c}$. This 
means that it represents a local divergence as $\epsilon \rightarrow 0$, and 
it can be therefore understood as the renormalization of a local operator in 
the action (\ref{szz}). 

Another important consequence of the 
actual expression of the functions $c_i (\lambda )$ is that
observable quantities derived from $Z_{\rm int }''$, as for instance the 
anomalous dimension $\gamma_c $ of the current operator, turn out to be 
finite in the limit $\epsilon \rightarrow 0$. The dimension $\gamma_c $ 
measures in particular the anomalous scaling of $\boldsymbol{\rho}_{c}$ 
under changes in the units of
energy and momentum in the system. The vertex $\boldsymbol{\Gamma}_{c}$
is formally a dimensionless quantity, but the renormalization process 
introduces scale dependence on the auxiliary momentum $\mu $, in such a way
that
\begin{equation}
\boldsymbol{\Gamma}_{c, {\rm ren}} \sim  \mu^{\gamma_c}
\end{equation}
The anomalous scaling of the vertex comes only from the dependence of the 
renormalization factor $Z_{\rm int }''$ on the $\mu $ scale, so that
\begin{equation}
\gamma_c = \frac{\mu }{Z_{\rm int }''}  \frac{\partial Z_{\rm int }''}{\partial \mu}
\label{ad}
\end{equation}

Assuming the general structure (\ref{poles}), it is in general a nontrivial
fact that the anomalous dimension $\gamma_c $ computed from $Z_{\rm int }''$ may
become finite in the limit $\epsilon \rightarrow 0$. The dependence on the scale
$\mu $ is encoded in the equation
\begin{equation}
\mu \frac{\partial \lambda }{\partial \mu} = -\epsilon \lambda
\label{rge}
\end{equation}
We can then express Eq. (\ref{ad}) in the form
\begin{eqnarray}
\gamma_c  & = &  \frac{\mu }{Z_{\rm int }''}  
           \frac{\partial \lambda}{\partial \mu}  
             \frac{\partial Z_{\rm int }''}{\partial \lambda}    \nonumber   \\
  & = &  - \frac{1}{Z_{\rm int }''} \: \lambda  
               \sum_{i=0}^{\infty} \frac{dc_{i+1}}{d\lambda } \frac{1}{\epsilon^i}
\end{eqnarray}
Alternatively, we can write the above equation as
\begin{equation}
\left( 1 + \sum_{i=1}^{\infty} \frac{c_i }{\epsilon^i} \right) \gamma_c
   =  - \lambda  \sum_{i=0}^{\infty} \frac{dc_{i+1}}{d\lambda }\frac{1}{\epsilon^i}
\label{alt}
\end{equation}
Assuming the finiteness of the anomalous dimension in the limit 
$\epsilon \rightarrow 0$, we get\cite{ram}
\begin{equation}
\gamma_c  =  -\lambda  \frac{d c_1}{d \lambda}
\end{equation}
and the consistency conditions for the cancellation of the poles at $\epsilon = 0$
\begin{equation}
\frac{d c_{i+1}}{d \lambda} = c_i \frac{d c_1}{d \lambda}
\label{cc}
\end{equation}

Quite remarkably, it can be seen that the expressions in (\ref{cs0}-\ref{cs}) 
satisfy identically the conditions (\ref{cc}). This holds for the functions 
$c_i (\lambda )$ which we have obtained analytically up to order $\lambda^7$. We 
have been also able to compute numerically their power series expansion up to 
order $\lambda^{18}$, checking that the equations (\ref{cc})
are verified order by order with the precision allowed by the calculation.
This provides a very strong evidence of the renormalizability of the theory,
implying that observable quantities like $\gamma_c $ can be computed from 
renormalized correlators to obtain finite results, dependent only on the
value of the physical coupling constant.

On the other hand, we note that the result for $\boldsymbol{\Gamma}_{c, {\rm ren}}$
is drastically modified when the electron self-energy corrections 
are included in the calculation of the vertex. As we have already seen in 
Sec. 3, the main effect of the electron self-energy is to renormalize the 
value of the Fermi velocity $v_F$. At the level discussed in that section, 
the self-energy corrections amount to perform the replacement in the inverse 
of the Dirac propagator
\begin{equation}
\gamma_0 \omega  - v_F \boldsymbol{\gamma} \cdot {\bf p} \; \rightarrow  \;
  \gamma_0 \omega  - v_F \boldsymbol{\gamma} \cdot {\bf p} - \Sigma_1 ({\bf p})
\end{equation}
with $\Sigma_1 ({\bf p})$ given by Eq. (\ref{loop}). It is then clear that
the electron self-energy diagrams can be incorporated to the ladder 
approximation encoded in Eq. (\ref{selfgc}) simply by trading the constant
$v_F$ by an effective Fermi velocity
\begin{equation}
\widetilde{v}_F({\bf p}) = v_F + \frac{e_0^2}{16 \pi^2} 
   (4\pi )^{\epsilon /2}   
  \frac{\Gamma \left(\tfrac{1}{2}\epsilon \right) 
              \Gamma \left(\tfrac{1-\epsilon}{2}\right) 
               \Gamma \left(\tfrac{3-\epsilon}{2}\right) }
  {  \Gamma (2 - \epsilon) }
   \frac{1}{|{\bf p}|^{\epsilon}}
\label{veff}
\end{equation}
The replacement of $v_F$ by $\widetilde{v}_F({\bf p})$ in
the formulas has the effect of iterating in the number of self-energy 
diagrams inserted in the electron and hole propagators building the vertex.
The electron self-energy corrections contribute therefore to supplement 
the ladder series previously considered, greatly improving the 
diagrammatic approach for the vertex $\boldsymbol{\Gamma}_{c}$.

In order to compute the renormalization factor $Z_{\rm int }''$ in the ladder
approximation with effective Fermi velocity $\widetilde{v}_F({\bf p})$, it is 
convenient to expand the factor $1/\widetilde{v}_F({\bf p})$ in powers of $e_0^2$ 
inside the integral of Eq. (\ref{integ}). 
The vertex $\boldsymbol{\Gamma}_{c}$ still
admits a solution like that in Eq. (\ref{exp}), as each order in the 
perturbative expansion can be represented in terms of the precedent by
integrals of the type shown in Eq. (\ref{integ}). The power series in $e_0^2$
contains now more poles in the $\epsilon $ parameter, as a result of the 
divergent behavior of $\widetilde{v}_F({\bf p})$ in Eq. (\ref{veff}). The poles
coming from the electron self-energy corrections can be however removed at once
by the renormalization of the Fermi velocity
\begin{equation}
v_F = Z_v v_{F, {\rm ren}}
\end{equation}
with
\begin{equation}
Z_v = 1 + b_1 \frac{1}{\epsilon}
\end{equation}
The coefficient needs simply to be adjusted to  
\begin{equation}
b_1 = - \frac{e^2}{16 \pi v_{F, {\rm ren}}}
\label{b1}
\end{equation}
leading then to a finite $\widetilde{v}_F({\bf p})$ written in terms of 
$v_{F, {\rm ren}}$.

We have again a general structure for the renormalization factor in this
improved approach
\begin{equation}
Z_{\rm int }'' = 1 + \sum_{i=1}^{\infty} \frac{\bar{c}_i (\lambda )}{\epsilon^i}
\label{polesb}
\end{equation}
The remarkable result is that, after writing the perturbative expansion for the 
vertex as a power series in the renormalized coupling
\begin{equation}
\lambda = \frac{e^2}{4\pi v_{F, {\rm ren}}}
\end{equation}
one needs just a simple first-order $1/\varepsilon $ term in (\ref{polesb}) to 
get rid of all the poles in $\boldsymbol{\Gamma}_{c}$. That is, the 
renormalized vertex becomes finite in the limit $\epsilon \rightarrow 0$ with 
the choice
\begin{eqnarray}
\bar{c}_1 (\lambda ) & = & - \frac{1}{4}  \lambda                                \\
\bar{c}_i (\lambda ) & = &   0    \;\;\;\;\;\;\;\;\;\;\;\;\;\;  i \geq 2
\end{eqnarray}

The simple pole structure of $Z_{\rm int }''$ in the improved ladder 
approximation implies the result
\begin{equation}
Z_v = Z_{\rm int }''
\label{ginv}
\end{equation}
We may see in this relation a nontrivial link between the renormalization of the 
$i v_F \boldsymbol{\gamma} \cdot \boldsymbol{\nabla}$ kinetic term and 
that of $\boldsymbol{\gamma} \cdot \boldsymbol{A}$ in the action (\ref{szz}). 
This feature points at a symmetry that is characteristic of a gauge invariant 
theory, and it can be explained in our case from inspection of the diagrams 
contributing to the self-energy discussed in Sec. 3 and to the vertices computed in 
this section, as we show next.

\subsection{No-renormalization of charge and current density operators}

The result (\ref{ginv}) implies, together with (\ref{zkin}), the preservation 
of the gauge invariance in the renormalized action (\ref{szz}). In this respect, 
there are actually Ward identities that can be derived from general principles,
relating different vertex functions of the theory. This is stressed for instance 
in Ref. \cite{juricic}, where it has been also emphasized the suitability of the 
dimensional regularization method to preserve the gauge symmetry of the theory.
We show here that the two relevant identities 
between the electron self-energy and the vertices $\Gamma_0$ and 
$\boldsymbol{\Gamma}_{c}$ can be easily obtained in the framework of the 
present many-body approach.

The main idea is that any electron self-energy correction can be converted 
into a contribution to the vertex $\Gamma_0$ or $\boldsymbol{\Gamma}_{c}$ by
taking the derivative with respect to the external frequency $\omega_k $ or 
the external momentum ${\bf k}$ of the self-energy. This fact relies on the 
expression of the derivatives of the free Dirac propagator
\begin{eqnarray}
\lefteqn{ \frac{\partial }{\partial \omega_k}  
  \frac{1}{\gamma_0 (\omega_k - \omega_p) - v_F \boldsymbol{\gamma} \cdot ({\bf k}-{\bf p})} }
                                                             \nonumber       \\
  &  &  \;\;\;\;\;\;\;\;\;\;\;   =  - \frac{1}
  {\gamma_0 (\omega_k - \omega_p) - v_F \boldsymbol{\gamma} \cdot ({\bf k}-{\bf p})} 
  \: \gamma_0  \: 
  \frac{1}
    {\gamma_0 (\omega_k - \omega_p) - v_F \boldsymbol{\gamma} \cdot ({\bf k}-{\bf p})}
                                                                                 \\
\lefteqn{  \frac{1}{v_F} \frac{\partial }{\partial {\bf k}}  
  \frac{1}{\gamma_0 (\omega_k - \omega_p) - v_F \boldsymbol{\gamma} \cdot ({\bf k}-{\bf p})} }
                                                               \nonumber        \\
  &  &   \;\;\;\;\;\;\;\;\;\;\;   =   \frac{1}
  {\gamma_0 (\omega_k - \omega_p) - v_F \boldsymbol{\gamma} \cdot ({\bf k}-{\bf p})} 
  \: \boldsymbol{\gamma} \: 
\frac{1}
  {\gamma_0 (\omega_k - \omega_p) - v_F \boldsymbol{\gamma} \cdot ({\bf k}-{\bf p})}
\end{eqnarray}
In any correction to the self-energy $\Sigma ({\bf k},\omega_k)$, one can 
choose the external frequency and momentum to circulate along the fermion 
lines that connect the two outer vertices of the diagram. Thus, taking
the derivative with respect to $\omega_k $ or ${\bf k}$ implies cutting any
of those internal lines in two pieces and inserting a vertex with the 
respective $\gamma_0$ or $\boldsymbol{\gamma}$ matrices. 

\begin{figure}

\vspace{0.5cm}

\begin{center}
\mbox{\epsfxsize 10cm \epsfbox{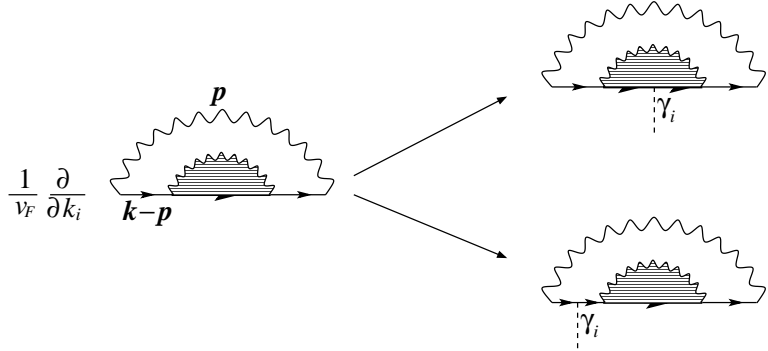}} 

\end{center}
\caption{Schematic representation of the Ward identity between the electron 
self-energy and the current density vertex in the ladder approximation.}
\label{five}
\end{figure}

The above construction becomes clear if one has in mind the diagrams 
building the self-consistent exchange approximation considered in Sec. 3. 
Taking the derivative of $\Sigma_{\rm SCEX} ({\bf k})$ with respect to 
$\omega_k $ gives identically zero, which is consistent with the absence of
corrections to $\Gamma_0$ in the ladder approximation (at zero momentum 
transfer). On the other hand, 
the derivative with respect to ${\bf k}$ gives rise to two different types
of diagrams contributing to $\boldsymbol{\Gamma}_{c}$, as shown in Fig. 
\ref{five}. Part of them corresponds to the kind of contributions that 
we were considering in the ladder approximation to $\boldsymbol{\Gamma}_{c}$
without electron self-energy corrections, as seen in the upper right of the 
figure. But the other part of the diagrams consists of vertex corrections with 
self-energy insertions in the electron and hole internal lines, as illustrated
in the lower right of Fig. \ref{five}. This shows that the differentiation of
$\Sigma_{\rm SCEX} ({\bf k})$ generates actually the whole set of vertex 
corrections in the ladder approximation supplemented with electron self-energy
corrections.

We can then write a Ward identity of the form
\begin{equation}
\frac{1}{v_F} \frac{\partial }{\partial {\bf k}} 
       \left( v_F  \boldsymbol{\gamma} \cdot {\bf k} + 
               \Sigma ({\bf k},\omega_k)  \right)  = 
   \boldsymbol{\Gamma}_{c}({\bf 0},0;{\bf k},\omega_k)
\label{wid}
\end{equation}
This identity implies that the renormalization of
the vertex $\boldsymbol{\Gamma}_{c}$ is dictated by that of the Fermi velocity
$v_F$. In this regard, the result $Z_v = Z_{\rm int }''$ found above becomes 
a natural consequence of Eq. (\ref{wid}).
Alternatively, these findings also stress the fact that the electron 
self-energy corrections cannot be neglected in a consistent approximation to 
the many-body theory of graphene, as they play a crucial role to build a gauge 
invariant effective action with the structure given by Eq. (\ref{szz}).

\section{Staggered (sublattice odd) charge density correlations}

We may also consider the renormalization of the staggered charge density 
operator antisymmetric under the exchange of the two sublattices of the 
graphene honeycomb lattice
\begin{equation}
\rho_3 ({\bf r}) = \overline{\psi}_i ({\bf r})  \psi_i ({\bf r})
\end{equation}
We will define the corresponding vertex by 
\begin{equation}
 \Gamma_3 ({\bf q},\omega_q;{\bf k},\omega_k)  =
   \langle  \rho_3 ({\bf q},\omega_q) 
        \psi_i ({\bf k}+{\bf q},\omega_k + \omega_q) 
           \overline{\psi}_i ({\bf k},\omega_k) \rangle_{\rm 1PI}
\label{gamma3}
\end{equation}
where 1PI denotes again that we take the irreducible part of the correlator.

The vertex $\Gamma_3 $ has a clear physical significance as it enters in the 
correlations of the staggered charge $\rho_3 ({\bf r})$, which is the order
parameter for chiral symmetry breaking in the many-body theory. A nonvanishing
expectation value $\langle \rho_3 ({\bf r}) \rangle \neq 0$ is the signal that 
a mass is dynamically generated for the Dirac fermions. This means that their 
hamiltonian gets effectively a term of the form
\begin{equation}
m \int d^2 r \; \overline{\psi}_i({\bf r})  \psi_i ({\bf r})
\end{equation}
With the mass term, the conduction and valence bands loose the perfect conical 
shape about the charge neutrality point, and a gap opens in the electronic 
spectrum. This trend of symmetry breaking is similar to that discussed long ago 
in Quantum Electrodynamics in two spatial dimensions\cite{pis,qed,kog,sem,sem2}. 
In the present context, the dynamical mass generation is also driven by the 
interaction, in such a way that the condensation of $\rho_3 ({\bf r})$ may proceed 
depending on the value of the coupling $e^2/4\pi v_F $ (and also on the number of 
fermion flavors, in a theory with a number $N$ of different fermion species).

\subsection{Staggered charge density vertex in ladder approximation}

We deal first with the vertex $\Gamma_3$ in the ladder approximation, which is
given again by the self-consistent equation represented diagrammatically in 
Fig. \ref{four}. As we are mainly interested in the cutoff dependence 
of the vertex, we can take in particular momentum transfer ${\bf q} = 0$ and 
$\omega_q = 0$. Given that $\Gamma_3$ must be anyhow proportional to the 
identity matrix, we have
\begin{align}
  \gamma_0 \frac{ -\gamma_0 \omega_p  + v_F \boldsymbol{\gamma} \cdot {\bf p} }
                      {-\omega_p^2  + v_F^2 {\bf p}^2 - i\eta } 
     \: \Gamma_3 ({\bf 0},0;{\bf p},\omega_p) \:
\frac{ -\gamma_0 \omega_p  + v_F \boldsymbol{\gamma} \cdot {\bf p} }
          {-\omega_p^2  + v_F^2 {\bf p}^2 - i\eta } \gamma_0    
  = -\frac{\Gamma_3 ({\bf 0},0;{\bf p},\omega_p)}{-\omega_p^2  + v_F^2 {\bf p}^2 - i\eta } 
\end{align}  
Thus, the self-consistent equation for the vertex in the ladder approximation
becomes
\begin{equation}
\Gamma_3 ({\bf 0},0;{\bf k},\omega_k) = 1 -  
    i \frac{e^2_0}{2}  \int \frac{d^D p}{(2\pi )^D} \frac{d\omega_p}{2\pi } 
      \Gamma_3 ({\bf 0},0;{\bf p},\omega_p) 
          \frac{1}{-\omega_p^2  + v_F^2 {\bf p}^2 - i\eta } 
               \frac{1}{|{\bf k}-{\bf p}|}
\label{selfg3}
\end{equation}
Eq. (\ref{selfg3}) can be further simplified by noticing that the solution
cannot depend on the frequency $\omega_k$. We end up then with the equation
\begin{equation}
\Gamma_3 ({\bf 0},0;{\bf k},\omega_k) = 1  +  \frac{e_0^2}{4} 
   \int \frac{d^D p}{(2\pi )^D} \Gamma_3 ({\bf 0},0;{\bf p},\omega_k) 
    \frac{1}{v_F |{\bf p}|} \frac{1}{|{\bf k}-{\bf p}|}
\label{selfcons}
\end{equation}

From dimensional arguments, the solution of Eq. (\ref{selfcons}) can be 
expressed in the form
\begin{equation}
\Gamma_3 ({\bf 0},0;{\bf k},\omega_k) = 
 1 + \sum_{n=1}^{\infty} \lambda_0^n 
                           \frac{s_n }{|{\bf k}|^{n\epsilon}} 
\label{ser}
\end{equation}
with $\lambda_0 = e_0^2/4\pi v_F $. Each term in the series (\ref{ser}) can be
obtained from the previous one, noticing that if we insert the general term
in the integral at the right-hand-side of Eq. (\ref{selfcons}) we get 
\begin{equation}
 \frac{e^2_0}{4} \int \frac{d^D p}{(2\pi )^D} \frac{1}{|{\bf p}|^{m\epsilon} }
            \frac{1}{v_F |{\bf p}|} \frac{1}{|{\bf k}-{\bf p}|} 
 =  \lambda_0 \frac{(4\pi )^{\epsilon /2}}{4}   
  \frac{\Gamma \left(\tfrac{m+1}{2} \epsilon  \right) \Gamma \left(\tfrac{1-(m+1)\epsilon}{2} \right) 
                                \Gamma \left(\tfrac{1-\epsilon}{2} \right) }
  { \sqrt{\pi} \Gamma \left(\tfrac{1+m\epsilon}{2} \right) \Gamma \left(1-\tfrac{m + 2}{2}\epsilon \right) }
 \frac{1}{|{\bf k}|^{(m+1)\epsilon}}  
\label{rec3}
\end{equation}
We observe that the result of the integral diverges in the limit 
$\epsilon \rightarrow 0$, leading to a sequence of higher-order poles in
the $\epsilon $ parameter as we look at higher perturbative levels in the 
solution (\ref{ser}).

The poles that appear in the computation of $\Gamma_3$ at $\epsilon = 0$ must
be reabsorbed by a suitable redefinition in the scale of the operator 
$\rho_3 ({\bf r})$. This is a composite field, susceptible of being renormalized
by a factor $Z_m$ which is independent of the renormalization of the elementary 
fields in the action\cite{amit}. This redefinition of $\rho_3 ({\bf r})$ 
translates into the multiplicative renormalization of the vertex
\begin{equation}
\Gamma_{3, {\rm ren}} = Z_m \Gamma_3
\label{mult3}
\end{equation}
The general structure of $Z_m$ must be
\begin{equation}
Z_m = 1 + \sum_{i=1}^{\infty} \frac{d_i (\lambda )}{\epsilon^i}
\end{equation}
in order to absorb all the poles generated by the recurrence relation 
(\ref{rec3}).

A nontrivial check of the renormalizability of the theory is that the vertex
$\Gamma_{3, {\rm ren}}$ must have a finite limit as $\epsilon \rightarrow 0$,
after making an appropriate choice of functions $d_i (\lambda )$ depending only
on the coupling constant $\lambda $. We have seen that this is the
case up to the order $\lambda^{24}$ we have pursued the numerical calculation
of $\Gamma_3$, finding a set of $d_i (\lambda )$ that do 
not depend on the momentum ${\bf k}$ of the vertex. The perturbative expansion 
of the functions can be computed analytically with some effort up to order 
$\lambda^8$, leading to
\begin{eqnarray}
d_1 (\lambda ) & = &  -\frac{1}{2} \lambda - \frac{1}{4} \log(2) \: \lambda^2 
           - \frac{1}{4} \log^2(2) \:  \lambda^3                           
        -  \frac{128 \log^3(2) + 3 \zeta(3)}{384} \:  \lambda^4  
                                                                \nonumber  \\
  &  &  - \frac{50 \log^4(2) +  3\log (2) \zeta(3)}{96} \: \lambda^5  
         -  \frac{4608 \log^5(2) + 480 \log^2(2) \zeta(3) 
             + 5 \zeta(5)}{5120}    \:  \lambda^6                \nonumber  \\
   &  &                                                 +  \ldots         \\
d_2 (\lambda ) & = &  \frac{1}{8} \: \lambda^2 + \frac{1}{8} \log(2) \: \lambda^3 
                  + \frac{5}{32} \log^2(2) \: \lambda^4    
         + \frac{176 \log^3(2) + 3 \zeta(3)}{768}  \: \lambda^5 
                                                                 \nonumber  \\
 &  & + \frac{192 \log^4(2) + 9 \log(2) \zeta(3)}{512} \: \lambda^6  +  \ldots    
                                                                           \\
d_3 (\lambda ) & = & -\frac{1}{48} \: \lambda^3 
      - \frac{1}{32} \log(2) \: \lambda^4 - \frac{3}{64} \log^2(2) \: \lambda^5 
 - \frac{464 \log^3(2) + 6 \zeta(3)}{6144} \: \lambda^6 + \ldots 
                                                                           \\
d_4 (\lambda ) & = &  \frac{1}{384} \: \lambda^4 
 + \frac{1}{192} \log(2) \: \lambda^5 + \frac{7}{768} \log^2(2) \: \lambda^6 
                                                     + \ldots               \\
d_5 (\lambda ) & = &  -\frac{1}{3840} \: \lambda^5 
              - \frac{1}{1536} \log(2) \: \lambda^6  + \ldots               \\
d_6 (\lambda ) & = &  \frac{1}{46080} \: \lambda^6   +  \ldots
\end{eqnarray}

An internal consistency check of the renormalizable theory is that, as in the
case of the vertex $\boldsymbol{\Gamma}_{c, {\rm ren}}$, the computation of
physical observables like the anomalous dimension $\gamma_m $ of the operator 
$\rho_3 $ has to provide a finite result in the limit 
$\epsilon \rightarrow 0$. $\gamma_m $ is defined in terms of the anomalous
scaling of the vertex as a function of the dimensionful parameter $\mu $ in
the renormalized theory,
\begin{equation}
\Gamma_{3, {\rm ren}} \sim  \mu^{\gamma_m}
\end{equation}
The dependence on $\mu $ arises from the renormalization factor $Z_m$, so
that
\begin{equation}
\gamma_m = \frac{\mu }{Z_m}  \frac{\partial Z_m}{\partial \mu}
\label{adm}
\end{equation}
We can follow the same derivation as in Eqs. (\ref{rge})-(\ref{alt}), ending
up in the equation
\begin{equation}
\left( 1 + \sum_{i=1}^{\infty} \frac{d_i (\lambda )}{\epsilon^i} \right) \gamma_m
   =  - \lambda  \sum_{i=0}^{\infty} \frac{d}{d\lambda }d_{i+1}(\lambda ) \: \frac{1}{\epsilon^i}
\label{altm}
\end{equation}
From Eq. (\ref{altm}) we obtain a finite answer at $\epsilon = 0$ for the 
anomalous dimension\cite{ram}
\begin{equation}
\gamma_m  =  -\lambda  \frac{d }{d \lambda}d_1(\lambda )
\label{anomm}
\end{equation}
provided that the recurrence relations
\begin{equation}
\frac{d }{d \lambda}d_{i+1}(\lambda ) = d_i (\lambda ) \: \frac{d }{d \lambda}d_1(\lambda )
\label{rr}
\end{equation}
are identically satisfied.

\begin{figure}

\vspace{0.5cm}

\begin{center}
\mbox{\epsfxsize 7cm \epsfbox{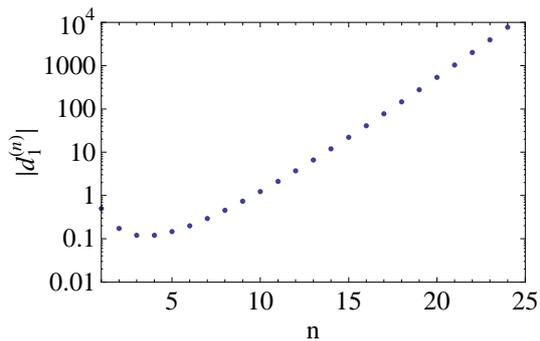}} 

\end{center}
\caption{Plot of the absolute value of the coefficients $d_1^{(n)}$ in the 
expansion of $d_1 (\lambda )$ as a power series of the renormalized coupling 
$\lambda $.}
\label{six}
\end{figure}

We have verified that the conditions (\ref{rr}) are fulfilled, up to the order 
$\lambda^{24} $ we have computed numerically the perturbative expansion of the
functions $d_i (\lambda )$. This means that the anomalous dimension $\gamma_m $
is perfectly well-defined by Eq. (\ref{anomm}) in the present framework. Regarding
the actual calculation, we have found that the perturbative expansion
\begin{equation}
d_1 (\lambda )  = \sum_{n=1}^{\infty}  d_1^{(n)} \lambda^n
\end{equation} 
behaves as a power series with a finite radius of 
convergence $\lambda_c$. The geometric growth of the coefficients $d_1^{(n)}$ 
is illustrated in the plot of Fig. \ref{six}. The radius of convergence can be 
obtained by computing the ratio between consecutive orders of $d_1^{(n)}$, and 
noticing that it converges towards a limit value $d$. It then turns out that
\begin{equation}
- d_1 (\lambda ) = \sum_{n=1}^{\infty} d^n \lambda^n \; + \; {\rm regular}  \;\;\; {\rm terms}
\end{equation} 
An excellent fit of the $n$-dependence of $d_1^{(n+1)}/d_1^{(n)}$ is achieved by
assuming the scaling behavior
\begin{equation}
\frac{d_1^{(n)}}{d_1^{(n-1)}} = d + \frac{d'}{n} + \frac{d''}{n^2} + \frac{d'''}{n^3} + \ldots
\end{equation} 
We obtain in this way an estimate of the radius of convergence
\begin{equation}
\lambda_c = \frac{1}{d} \approx 0.456947
\label{crit}
\end{equation}

The singular behavior of the anomalous dimension $\gamma_m $ is the 
manifestation of the divergence of the vertex $\Gamma_{3, {\rm ren}}$ at the
critical coupling $\lambda_c$. It also implies the divergence of the 
correlators of the staggered charge density $\rho_3 ({\bf r})$ as these
need to be multiplicatively renormalized by factors of $Z_m$, which means
that their anomalous dimensions are given by multiples of $\gamma_m $.
What we find therefore at the critical point $\lambda_c$ is the signature
of the dynamical breakdown of the chiral symmetry in the Dirac theory. The 
predicted critical coupling in Eq. (\ref{crit}) turns out to be a very accurate
approximation to the critical value
\begin{equation}
\lambda_c = \frac{8\pi^2}{\left(\Gamma \left( \tfrac{1}{4} \right) \right)^4}
\end{equation}
which has been obtained by Gamayun {\em et al.} in Ref. \cite{gama} by a quite
different approach, consisting in the self-consistent resolution of the gap 
equation for the Dirac fermions (for another connection between the branching 
point of the gap equation and the singularity in the vertex, see also  
Ref. \cite{juan}). This remarkable coincidence between 
the results of two completely different methodologies can be taken as the 
reflection that they are encoding at the end an equivalent sum of many-body 
corrections, thus providing a nice check of the reliability of our computational 
framework based on the ladder approximation to vertex functions.

\subsection{Staggered charge density vertex supplemented with self-energy corrections}

We have anyhow to keep in mind that the electron self-energy corrections need 
to be incorporated to reach sensible results for $\Gamma_{3, {\rm ren}}$, as we
learned from the renormalization of the current density vertex. In the present
case, the ladder approximation can be also improved by inserting the series of 
electron self-energy diagrams in the internal electron and hole states of the 
vertex. From a computational point of view, this can be achieved by 
replacing the constant $v_F$ in the integrand of Eq. (\ref{selfcons}) by the
effective Fermi velocity $\widetilde{v}_F({\bf p})$ dressed with the 
self-energy corrections in Eq. (\ref{veff}). It can be easily seen that 
expanding the latter in powers of $e_0^2$ corresponds to generating the 
iteration of self-energy corrections to the internal fermion 
propagators in the equation of Fig. \ref{four}.

A solution of the form (\ref{ser}) can be still found for 
$\Gamma_3 ({\bf 0},0;{\bf k},\omega_k)$, where now each term in the series
can be obtained from all the precedent by expanding $\widetilde{v}_F({\bf p})$
in powers of $e_0^2$ in Eq. (\ref{selfcons}) and using repeatedly the formula 
(\ref{rec3}). The Fermi velocity $v_F$ needs to be renormalized to absorb the
divergence of $\widetilde{v}_F({\bf p})$ in the limit $\epsilon \rightarrow 0$,
for which we define 
\begin{equation}
v_F = Z_v v_{F, {\rm ren}}
\label{vren}
\end{equation}
As in subsection 4.2, $Z_v$ just contains a simple pole
\begin{equation}
Z_v = 1 + b_1 \frac{1}{\epsilon}
\end{equation}
with $b_1 = -e^2/16\pi v_{F, {\rm ren}}$.
After subtraction of the self-energy pole, 
the rest of poles in $\Gamma_3$ must be reabsorbed by the multiplicative
renormalization (\ref{mult3}), where now $Z_m$ is given by a different series
\begin{equation}
Z_m = 1 + \sum_{i=1}^{\infty} \frac{\tilde{d}_i (\lambda )}{\epsilon^i}
\end{equation}

One can see that in this case $\Gamma_{3, {\rm ren}}$ can be made also finite
in the limit $\epsilon \rightarrow 0$, with a set of functions 
$\tilde{d}_i (\lambda )$ that only depend on the renormalized coupling constant
\begin{equation}
\lambda = \mu^{-\epsilon} Z_v \lambda_0 = \frac{e^2}{4\pi v_{F, {\rm ren}}}
\label{lren}
\end{equation}
The first orders in the perturbative expansion are given for instance by
\begin{eqnarray}
\tilde{d}_1 (\lambda )  & = &   - \frac{1}{2} \lambda - \frac{1}{8} \log(2) \: \lambda^2 
      - \frac{\pi ^2 + 120 \log ^2(2)}{1152}   \lambda^3               
        -  \frac{10 \pi ^2 \log (2)+688 \log ^3(2)+15 \zeta (3)}{6144}   \lambda^4   
                                                               \nonumber  \\
 & &  - \frac{13 \pi ^4+2064 \pi ^2 \log ^2(2)+144 \left(716 \log ^4(2)+37 \log (2) \zeta (3)\right)}{737280} \: \lambda^5   
                                                                             \nonumber \\
 & &  - \frac{1}{{2949120}} \left(169 \pi^4 \log(2) + 567744 \log^5(2) + 49320 \log^2(2) \zeta(3)  \right.
                                                               \nonumber    \\
 & &  \;\;\;\;\;\;\;\;\;\;  \;\;\;\;\;\;\;\;\;\;
   \left.  + 5 \pi ^2 \left(2864 \log^3(2) + 37 \zeta(3)\right) + 1125 \zeta(5) \right) \: \lambda^6  
                                            +    \ldots                       \\
\tilde{d}_2 (\lambda ) & = &  \frac{1}{16} \: \lambda^2 + 
            \frac{1}{24} \log(2) \: \lambda^3                         
         + \frac{5 \pi ^2 + 744 \log ^2(2)}{18432}  \lambda^4          \nonumber  \\      
 & &       + \frac{110 \pi ^2 \log (2)+8592 \log ^3(2)+135 \zeta (3)}{184320}  \: \lambda^5 
                                                              \nonumber     \\
  & &  + \frac{293 \pi^4 + 58944 \pi^2 \log^2(2) + 72 \left(44392 \log^4(2) + 1779 \log(2) \zeta(3)\right)}{53084160}
                 \: \lambda^6                    +  \ldots                     \\
\tilde{d}_3 (\lambda ) & = & - \frac{1}{768} \log (2) \: \lambda^4 
           - \frac{\pi ^2+360 \log ^2(2)}{184320} \lambda^5           \nonumber   \\
 & &   -  \frac{100 \pi^2 \log(2) + 11904 \log^3(2) + 45 \zeta(3)}{4423680}  \:  \lambda^6
            + \ldots                                                            \\
\tilde{d}_4 (\lambda ) & = & - \frac{1}{7680} \log (2) \: \lambda^5   
         - \frac{\pi^2 + 280 \log^2(2)}{1474560}  \:  \lambda^6    +  \ldots        \\
\tilde{d}_5 (\lambda ) & = & - \frac{1}{61440} \log(2) \: \lambda^6    +  \ldots 
\label{coeff}
\end{eqnarray}
We have computed the expansions of the functions $\tilde{d}_i (\lambda )$
numerically up to order $\lambda^{24}$, checking that the coefficients do not 
depend on the momentum ${\bf k}$ of the vertex. This is the essential 
requirement guaranteeing the renormalizability of the theory, by which the
divergences in the limit $\epsilon \rightarrow 0$ can be absorbed into the 
redefinition of a finite number of local operators.

We can proceed to the computation of the anomalous dimension $\gamma_m $ in
the present approach, taking again into account that the scale dependence of
$\Gamma_{3, {\rm ren}}$ stems from $Z_m$, so that
\begin{equation}
\gamma_m   =   \frac{\mu }{Z_m}  
           \frac{\partial \lambda}{\partial \mu}  
             \frac{\partial Z_m}{\partial \lambda}
\label{chr}
\end{equation}
The dependence of the renormalized coupling $\lambda $ on $\mu $ can be 
obtained by differentiating (\ref{lren}), leading to
\begin{equation}
\mu \frac{\partial \lambda}{\partial \mu } = -\epsilon \lambda + 
   \frac{\lambda }{Z_v} \mu \frac{\partial \lambda}{\partial \mu } \frac{\partial Z_v}{\partial \lambda }
\end{equation}
Using the fact that $Z_v = 1 + \lambda (\partial Z_v / \partial \lambda )$, we obtain
\begin{eqnarray}
\mu \frac{\partial \lambda}{\partial \mu }  & = & -\epsilon \lambda \: Z_v        \nonumber   \\
   & = &   -\epsilon \lambda  - \lambda \: b_1  (\lambda )
\end{eqnarray}
This result can be now introduced in Eq. (\ref{chr}), finding that
\begin{equation}
\gamma_m = \frac{1}{Z_m} (-\epsilon \lambda  - \lambda \: b_1 (\lambda) )
     \sum_{i=1}^{\infty }  \frac{1}{\epsilon^i} \frac{d}{d\lambda } \tilde{d}_i (\lambda)
\end{equation}
Equivalently, we can write
\begin{equation}
\left( 1 + \sum_{i=1}^{\infty } \frac{\tilde{d}_i (\lambda)}{\epsilon^i} \right)  \gamma_m 
   =  - \lambda \sum_{i=0}^{\infty }  \frac{1}{\epsilon^i} \frac{d}{d\lambda } \tilde{d}_{i+1} (\lambda)
      - \lambda b_1 (\lambda) 
      \sum_{i=1}^{\infty }  \frac{1}{\epsilon^i} \frac{d}{d\lambda } \tilde{d}_i (\lambda)
\end{equation}
Assuming that $\gamma_m $ must have a finite limit as $\epsilon \rightarrow 0$,
we get
\begin{equation}
\gamma_m = -\lambda \frac{d}{d\lambda } \tilde{d}_1 (\lambda)
\label{gamm}
\end{equation}
together with the conditions for the cancellation of all the pole contributions 
to $\gamma_m $\cite{ram} 
\begin{equation}
 \tilde{d}_i (\lambda) \: \frac{d}{d\lambda } \tilde{d}_1 (\lambda) =
\frac{d}{d\lambda } \tilde{d}_{i+1} (\lambda) + b_1 (\lambda) \: 
 \frac{d}{d\lambda } \tilde{d}_i (\lambda)
\label{rem}
\end{equation}

\begin{figure}

\vspace{0.5cm}

\begin{center}
\mbox{\epsfxsize 7cm \epsfbox{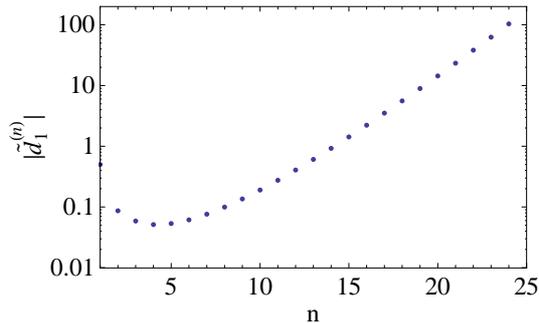}} 

\end{center}
\caption{Plot of the absolute value of the coefficients $\tilde{d}_1^{(n)}$ in the 
expansion of $\tilde{d}_1 (\lambda )$ as a power series of the renormalized 
coupling $\lambda $.}
\label{seven}
\end{figure}

Quite remarkably, we have verified that the conditions (\ref{rem}) are 
indeed satisfied by the perturbative series of the functions 
$\tilde{d}_i (\lambda )$, at least up to the order $\lambda^{24}$ we have been 
able to carry out numerically the expansion. This guarantees that Eq. (\ref{gamm})
can be used to obtain a finite result for $\gamma_m $, only dependent on the
value of the coupling constant $\lambda $. Computing the perturbative expansion
\begin{equation}
\tilde{d}_1 (\lambda )  = \sum_{n=1}^{\infty}  \tilde{d}_1^{(n)} \lambda^n
\end{equation}
we have checked that in this case again the coefficients $\tilde{d}_1^{(n)}$ grow
geometrically with the order $n$, as shown in Fig. \ref{seven}. 
The ratio $\tilde{d}_1^{(n+1)}/\tilde{d}_1^{(n)}$
converges to a limit value $\tilde{d}$, in such a way that
\begin{equation}
- \tilde{d}_1 (\lambda ) = \sum_{n=1}^{\infty} \tilde{d}^n \lambda^n \; + \; {\rm regular}  \;\;\; {\rm terms}
\end{equation}
We have found that the points $\tilde{d}_1^{(n+1)}/\tilde{d}_1^{(n)}$ can be fitted
quite accurately by the scaling behavior
\begin{equation}
\frac{\tilde{d}_1^{(n)}}{\tilde{d}_1^{(n-1)}} = \tilde{d} + \frac{\tilde{d}'}{n} 
   + \frac{\tilde{d}''}{n^2} + \frac{\tilde{d}'''}{n^3} + \ldots
\end{equation} 
We obtain in this way a finite radius of convergence for the perturbative 
expansion 
\begin{equation}
\lambda_c = \frac{1}{\tilde{d}}  \approx 0.544775
\label{crit2}
\end{equation}
where the error is estimated to be in the last digit (in similar fashion as for 
the critical value given in (\ref{crit}). This represents a considerable gain
in precision with respect to the calculation reported in Ref. \cite{prb}, which
is now made possible by our ability to carry out the power series expansion for 
$\tilde{d}_1 (\lambda )$ up to seven orders beyond that achieved in that paper.

The existence of a critical coupling $\lambda_c $ implies that the many-body 
theory of interacting Dirac fermions enters a new phase for sufficiently 
large strength of the Coulomb interaction. According to our preceding 
discussion, this phase is characterized by the condensation of the staggered
charge density, which signals the dynamical breakdown of the chiral symmetry
of the theory. We find therefore that this phenomenon takes place even after
accounting for the electron self-energy corrections in the many-body theory. 
Thus, the effect of Fermi velocity renormalization tends to reduce the 
effective interaction strength, but it does not prevent the dynamical mass
generation, leading instead to a larger value (\ref{crit2}) of the critical 
coupling in comparison to the approach without switching on the electron 
self-energy corrections (see also Ref. \cite{sabio}).

\section{Conclusions}

The above results provide strong evidence that the model of
interacting Dirac fermions in graphene constitutes a completely 
renormalizable field theory, in the sense that all the cutoff dependences 
can be absorbed into the redefinition of a finite number of local operators.
We have proven this fact for different vertex functions in the ladder 
approximation, as well as when this is supplemented by electron self-energy 
corrections to the electron and hole states in the vertices. We have seen 
that there is a nontrivial cancellation of poles in the dimensionally regularized
anomalous dimensions of different fermion bilinears, a feature typically 
enforced by nonperturbative equations for the residues of the poles in
renormalizable quantum field theories. It would be then interesting 
to pursue the program of renormalization in other approximations with higher 
diagrammatic content, testing the cutoff independence of 
anomalous exponents in the graphene electron system.

The other important conclusion we reach is that the incorporation of
electron self-energy corrections is in general 
required to preserve the gauge invariance that relates the kinetic and the 
interaction terms in the effective action of the theory. When looking at 
corrections to the current density vertex in the ladder approximation, gauge 
invariance implies in particular the complete cancellation
between the cutoff-dependent part of self-energy and vertex corrections to
all orders in the perturbation expansion,
as we have checked explicitly in our calculation with dimensional 
regularization. 

From a practical point of view, our computational approach has allowed us
to address the question of the chiral symmetry breaking in the interacting 
theory of Dirac fermions in graphene. Making use of the renormalizability
of the theory, we have shown that quantities like the anomalous dimension 
of the staggered charge density operator can be determined in terms of the 
renormalized coupling alone, allowing to characterize the dynamical symmetry 
breaking at the critical coupling given by the finite radius of convergence 
of the perturbative expansion.

We note anyhow that the critical value found for the coupling 
$\lambda $ cannot be used directly to 
predict the onset of dynamical symmetry breaking in real graphene, as the 
$e$-$e$ interaction is affected in general by screening processes that may reduce 
significantly its strength. This means that the critical coupling $\lambda_c $
we have computed should be referred actually to the effective interaction strength 
after incorporating screening corrections. These depend in general on 
intrinsic factors, like the number $N$ of different fermion flavors that enter 
in the polarization of the system. Under static RPA screening, for instance, the
coupling $\lambda $ of the effective interaction is related to the bare fine
structure constant $\alpha $ in graphene by the expression
\begin{equation}
\lambda = \frac{\alpha }{1+ \frac{N\pi}{4}\alpha }
\end{equation}
$N$ being the number of four-component Dirac fermions. In the physical case
$N = 2$, the nominal coupling of graphene in vacuum $\alpha \approx 2.2$
leads to the estimate $\lambda \approx 0.49$, which is above the critical 
coupling (\ref{crit}) obtained in the pure ladder approximation, but below  
the value (\ref{crit2}) found after incorporating the electron self-energy 
corrections. The latter critical value of $\lambda $ translates 
therefore into a more stringent bound on the nominal value of $e^2/v_F$ for the 
development of chiral symmetry breaking. It is worth noticing however that the
more sensible screening approaches, considering the dynamical
polarization of electron-hole pairs, still predict the dynamical mass 
generation for the coupling $\alpha \approx 2.2$ corresponding to graphene 
isolated in vacuum\cite{ggg,prb}.

The reliability of our renormalization procedure is reinforced by the
fact that the critical coupling we obtain from the sum of ladder diagrams
for the vertex
matches with great accuracy the value found within a quite different approach 
to chiral symmetry breaking in graphene, based on the self-consistent resolution
of the gap equation for the Dirac fermions\cite{gama}. This
provides good motivation to check whether a similar agreement can be reached
after incorporating the electron self-energy corrections in the self-consistent
gap equation, as well as to extend by any other feasible means the approach 
devised in the present paper.

\section{Acknowledgments}

The financial support from MICINN (Spain) through grant
FIS2008-00124/FIS is gratefully acknowledged. This research was also supported 
in part by the National Science Foundation under Grant No. NSF PHY11-25915.
We also thank the hospitality of the Kavli Institute for Theoretical Physics
(Santa Barbara), where this work has been completed.



\end{document}